\documentstyle[12pt,psfig,epsfig]{article}
\newcommand{\beq}{\begin{equation}}
\newcommand{\eeq}{\end{equation}}
\newcommand{\beqarray}{\begin{eqnarray}}
\newcommand{\eeqarray}{\end{eqnarray}}
\def\lsim{\raise0.3ex\hbox{$\;<$\kern-0.75em\raise-1.1ex\hbox{$\sim\;$}}}
\def\gsim{\raise0.3ex\hbox{$\;>$\kern-0.75em\raise-1.1ex\hbox{$\sim\;$}}}
\def\para{\vspace{0.3cm}\noindent}

\def\gev{\,{\rm GeV}}

\begin{document}
\begin{center}
{\large \bf Ultrahigh Energy Cosmic Rays from Gamma Ray Bursts: Implications of the Recent Observational Results by Milagro}

\medskip

{Nayantara Gupta \footnote{Email address: nayan@imsc.res.in}}\\  
{\it Institute of Mathematical Sciences,\\
 C.I.T Campus, Taramani, Chennai 600113\\
 INDIA.}  
\end{center}

\begin{abstract}
It has been speculated earlier that Gamma Ray Bursts are sources of ultrahigh energy cosmic rays. Recently, the search for high energy photons from Gamma Ray Bursts by Milagro group has put limits on the isotropic luminosity of these transient sources in very high energy photons. The implications of the results obtained by Milagro to our understanding of the ultrahigh energy cosmic ray spectrum from these sources have been discussed in the present work.
\end{abstract} 

\vskip 2cm
Gamma Ray Bursts (GRBs) are likely to be excellent sites for particle acceleration and are expected to emit ultrahigh energy cosmic rays (UHECR)
\cite{vietri1, waxman1}. 
Earlier, several groups have investigated the possibility of $TeV$ photon emission from GRBs \cite{totani,pijush} in the burst phase. Mechanisms have been invoked which can give rise to $TeV$ energy photons inside GRBs. They are $(i)$ synchrotron self-compton (SSC) process, which involves inverse compton scattering of sufficiently high energy electrons on the ambient synchrotron photons within the source, the synchrotron photons being radiated by non-thermal population of electrons inside the GRB, $(ii)$ photo-pion production by ultrahigh energy protons within the source and the subsequent decay of the neutral pions to photons, $(iii)$ synchrotron radiation by ultrahigh energy protons having energies above $10^{20}eV$ in the magnetic field within the source. $TeV$ photons would interact with the low energy photons inside the source and also with the photons of the intergalactic medium. Their flux would be diminished due to their annihilation with the low energy photons before reaching the Earth.

\para

Milagro telescope located in northern New Mexico observes the northern sky for $100 GeV$ to $100 TeV$ $\gamma$-ray emission. The Milagro group searched for
$40s - 3h$ transient very high energy emission \cite{milagro1,milagro2}. They have obtained flux limits and also constrained the absolute luminosity of GRBs in very high energy $\gamma$-rays. Their limits on very high energy $\gamma$-ray luminosity of GRBs is necessarily model dependent. The following assumptions have been made in their analysis: $(i)$ the photon spectrum is proportional to $E_{\gamma}^{-2}$ in the energy range $100 GeV$ to $21 TeV$, $(ii)$ GRBs are isotropically distributed in the Universe, $(iii)$ GRBs follow star formation rate in $z$ and $(iv)$ absorption of TeV photons by extragalactic background light has been included, following \cite{bull}. 
Assuming that all GRBs are emitting a $80 sec$ pulse of very high energy photons, they have found that for an average distribution of $4.8 GRBs/Gpc^3/year$ over $0<z<0.5$ the $90\%$ confidence upper limit on their isotropic lumisoty is $10^{51} erg/sec$ and for an average of $0.8 GRBs/Gpc^3/year$ over $0<z<0.5$ the corresponding upper limit on their luminosity being $10^{52}erg/sec$.
These limits on luminosity of GRBs also depend on the duration of the bursts.
Milagro group has used duration of bursts as $``80 sec"$ as a representative value in their work \cite{milagro2} to illustrate how an upper limit can be set on the high energy photon luminosity of GRBs.

\para

The physical mechanism of very high energy photon production inside GRBs may put  constraints on the expected UHECR spectrum of protons from these sources. The SSC model of high energy photon production does not involve ultrahigh energy cosmic ray nuclei and this mechanism has no role in constraining the expected UHECR spectrum from GRBs.
In the other two models protons are accelerated by internal shocks in the magnetic field inside a GRB. The resulting ultrahigh energy proton spectrum is expected to be a power law as given below. 
\beq
\frac{dn_p(E_p)}{dE_p}=A_p\left\{ 
\begin{array}{ll}E_p^{-\alpha}\,, & \mbox{if 
$E_p < E_{pb}$}\,,\\
E_{pb}E_p^{-\alpha-1}\,, & \mbox{if $E_p \geq E_{pb}$\,.}
\end{array}
\right.\label{A_p_2}
\eeq
$A_p$ is the normalisation constant and spectral index $\alpha=2\pm 0.2$ \cite{hillas}. $E_{pb}$ is the break energy in the proton spectrum of a GRB.
The protons accelerated in an irregular magnetic field inside the GRB are emitting synchrotron photons. In the fireball model of a GRB, expanding ultrarelativistic shells of matter or wind moving with different velocities collide with each other and also with the external medium. If the expansion time of the ultrarelativistic shell is longer than the synchrotron energy loss time by protons then the protons would loss all their energies by synchrotron photon emission, otherwise they would  lose only a fraction of their total energies during the expansion of the ultrarelativistic wind. The break energy of the proton spectrum $E_{pb}$ represents the energy above which the protons would lose all their energies by synchrotron radiation within one expansion time scale of the wind.

\para

Some of the ultrahigh energy protons may come out from the GRBs and give rise to the UHECR spectrum observed on Earth \cite{agasa1,hires,hav}.
For $\alpha=2$, the resulting high energy $\gamma$-ray spectrum in the photo-pion production model is expected to be proportional to $E_{\gamma}^{-1}$ in the energy range of our interest. 
The photo-pion production scenario is not an efficient mechanism of very high energy $\gamma$-ray production \cite{totani}, hence it has not been included 
in the present discussion.
In the proton-synchrotron model of very high energy $\gamma$-ray production the $\gamma$-ray spectrum is proportional to $E_{\gamma}^{-1.5}$ and $E_{\gamma}^{-2}$ below and above the photon spectrum break energy respectively for $\alpha=2$ \cite{totani,pijush}.
Using the proton-synchrotron mechanism of very high energy photon production 
 and the limit obtained by Milagro group \cite{milagro2} the contribution of GRBs to the UHECR spectrum observed on Earth above energy $10^{19} eV$ has been explored in the present work. 

\vskip 0.5cm
{\it The Proton-Synchrotron Mechanism of Very High Energy Photon Production:}
\vskip 0.5cm
  This mechanism has been discussed in detail earlier in these two papers \cite{totani,pijush}. GRBs emit $\gamma$-rays, X-rays and also photons in optical and radio frequencies. KeV-MeV energy photons inside a GRB are likely to be produced by the synchrotron radiation of shock accelerated electrons. The energy density of these photons (KeV-MeV) in the wind rest frame, $U_{\gamma}$, would be almost equal to the energy density of the electrons $U_e$,  if electron synchrotron emission is an efficient mechanism  inside the fireball. Initially, the accelerated protons carry $m_p/m_e$ times more energy than the accelerated electrons. If we denote the energy density of the accelerated protons by $U_p$ then $U_p=\frac {m_p}{m_e} U_e \sim \frac {m_p}{m_e} U_{\gamma}$.
In this model it is assumed that the magnetic field inside the fireball
 has an energy density almost equal to the total energy density of the fireball. The total energy density $U$ is nearly equal to $U_p$ as most of the energy is carried by the shock accelerated protons inside the fireball. 
One can relate $U_B$ and $U_p$ from the above discussion in the following way,
\beq
U_{B}=\xi_{B}U_{p}\,,\label{U_b}
\eeq
$\xi_{B}$ is the equipartition parameter whose value is of the order of unity.
The magnetic field energy density can be expressed as,
\beq
\frac{B^2}{8\pi}=\xi_{B}\frac{m_p}{m_e}U_{\gamma}\,,\label{Bden}
\eeq
The energy density of the low energy photons $U_{\gamma}$ in the wind rest frame can be expressed in terms of the luminosity of these photons $L_{L{\gamma}}$ in that frame. The low energy photon luminosity of a GRB in the wind rest frame is related to its observed luminosity on Earth $L_{L{\gamma}}^{ob}$ by the factor $(1+z)^2$ which takes into account the corrections in time and energy due to the redshift ($z$) of the GRB. If the GRB has an average Lorentz factor $\Gamma$ and the time interval between two successive emission of ultrarelativistic shells (variability time scale) as measured by the observer on Earth is $\triangle t^{ob}$, then the observed luminosity of that GRB in low energy photons $L_{L{\gamma}}^{ob}$ can be expressed as a function of the parameters $z$, $\Gamma$, $\triangle t^{ob}$ and energy density $U_{\gamma}$. The light curve of a GRB  usually shows variation of the intrinsic variability time scale in a wide range of values and the intrinsic variability time of a GRB is much less than the total duration of the burst. The proton-synchrotron mechanism of high energy photon production is not significantly sensitive to the variation in $\triangle t^{ob}$ over the range $0.1sec \le \triangle t^{ob} \le 10 sec$ \cite{totani}. The expanding ultrarelativistic shells of matter collide with each other and their energies are dissipated. The characteristic distance, at which they collide and internal shocks are formed, is called the dissipation radius. The dissipation radius as measured at the source in the observer's reference frame has been denoted by $r_{d}^{s}$. We would refer to the observer's reference frame at the  source as the source rest frame where the ultrarelativistic wind moves with Lorentz factor $\Gamma$.  
\beq
L_{L{\gamma}}^{ob}=\frac{L_{L{\gamma}}}{(1+z)^2}=\frac{1}{(1+z)^2}4\pi(r_{d}^{s})^2{\Gamma}^2cU_{\gamma}\,,\label{lum_low}
\eeq
where, 
\beq
r_d^{s}=2\Gamma^2c\triangle t^{ob}/(1+z)\,,\label{diss_rad}
\eeq
The magnetic field $B$ inside the GRB can be expressed in terms of the GRB parameters using equations (\ref{Bden})-(\ref{diss_rad}).  
The shock accelerated protons inside the fireball would emit high energy photons by synchrotron radiation during the expansion of the ultrarelativistic shells of matter or wind. The characteristic energy of these high energy photons $E_{\gamma}$ depends on the value of the proton energy $E_p$ and the magnetic field $B$.
\beq
E_{\gamma}=\frac{3}{2}\frac{ehBE_p^2}{2\pi m_p^3 c^5}<sin \theta>\,,\label{E_g}
\eeq
where, $e$ is the charge of the proton and the average value of the sine of the pitch angle $\theta$ of the proton is $\pi/4$. 
The proton energy and synchrotron radiated photon energy can be related as described below using equation (\ref{E_g}),
\beq
E_p(GeV)=\kappa_1 (E_{\gamma}(GeV))^{1/2}\,,\label{E_p_w}
\eeq

After expressing $B$ in terms of the GRB parameters we get in the wind rest frame 

\beq
\kappa_1=1.3\times10^{8}\frac{\Gamma_{300}^{3/2} (\frac{\triangle t^{ob}}{sec})^{1/2}}{\xi_B^{1/4}(L_{L\gamma,51}^{ob})^{1/4} (1+z)}\,,\label{k_1}
\eeq

where, $\Gamma=\Gamma_{300}\times 300$ and observed luminosity in $KeV-MeV$ photons $L_{L\gamma}^{ob}=L_{L\gamma,51}^{ob}\times 10^{51} erg/sec$.
The energies of any particle in the observer's rest frame and in the wind rest frame are related as follows
\beq
E^{ob}=\frac{\Gamma}{1+z}E\,,\label{ref_frame}
\eeq
This general relation in equation (\ref{ref_frame}) can be substituted in equation (\ref{E_p_w}) to relate the proton energy and synchrotron photon energy in the observer's rest frame on Earth.
\beq
E_p^{ob}(GeV)=\kappa_2 (E_{\gamma}^{ob}(GeV))^{1/2}\,,\label{E_p_ob}
\eeq
where,
\beq
\kappa_2=2.3\times10^{9}\frac{\Gamma_{300}^2 (\frac{\triangle t^{ob}}{sec})^{1/2}}{\xi_B^{1/4}(L_{L\gamma,51}^{ob})^{1/4} (1+z)^{3/2}} \,,\label{k_2}
\eeq

The fractional energy loss of the protons by synchrotron radiation during one expansion time scale ($t_{exp}$) of the ultrarelativistic wind is as follows
\beq
f_{ps}(E_p)=\frac{t_{exp}}{t_{ps}(E_p)}\,,\label{f_ps}
\eeq
where, the inverse of the proton synchrotron energy loss time scale is
\beq
\frac{1}{t_{ps}(E_p)}\equiv\frac{1}{E_p}\left(\frac{dE}{dt}\right)_{ps}(E_p)\,,\label{t_ps}
\eeq
Assuming, $E_p>>m_p c^2$ and isotropic distribution of the pitch angles the right side of the above equation can be expressed as
\beq
\frac{1}{E_p}\left(\frac{dE}{dt}\right)_{ps}(E_p)\simeq c\frac{4}{9E_p}\left(\frac{e^2}{m_pc^2}\right)^2\left(\frac{E_p}{m_p c^2}\right)^2 B^2\,,\label{E_ps}
\eeq
The expansion time scale is related to the dissipation radius $r_d^s$ and the Lorentz factor $\Gamma$.
\beq
t_{exp}\simeq \frac{1}{\Gamma}\frac{r_d^s}{c}\,,\label{t_exp}
\eeq
The break energy in the ultrahigh energy proton spectrum can be obtained by
 equating the two time scales $t_{ps}(E_p)$ and $t_{exp}$.
\beq
t_{ps}(E_p=E_{pb})=t_{exp}\,,\label{E_pb_cond}
\eeq
 Above the break energy $E_{pb}$ the protons lose all their energies by synchrotron radiation during the expansion of the ultrarelativistic wind hence, the
fractional energy loss by protons can be expressed as
\beq
f_{\rm ps}(E_p)=\left\{ 
\begin{array}{ll}\frac{E_p}{E_{pb}}\,, & \mbox{if 
$E_p < E_{pb}$}\,,\\
1\,, & \mbox{if $E_p \geq E_{pb}$\,.}
\end{array}
\right.\label{f_p_2}
\eeq
The expansion time scale $t_{exp}$ of the wind and synchrotron energy loss time scale of protons $t_{ps}(E_p)$ can be expressed in terms of the variability time scale $\triangle t^{ob}$, redshift $z$ and Lorentz factor $\Gamma$ of the GRB. Using equation (\ref{E_pb_cond}) one can obtain the expression for the break energy in the proton spectrum in the wind rest frame as given below
\beq
E_{pb}\simeq 1.6\times 10^8 \frac{\Gamma^5_{300}\left(\frac {\triangle t^{ob}}{sec}\right)}{(1+z)^3\xi_B\left(L^{ob}_{L,51}\right)}GeV\,,\label{E_pb}
\eeq
The break energy in the synchrotron photon spectrum due to the break energy in the proton spectrum appears at $E_{{\gamma}b}$.
\beqarray
E_{\gamma b} & \equiv & \kappa_1^{-2}
\left(\frac{E_{pb}}{\gev}\right)^2 GeV,\\
 & = & 1.5
\frac{\Gamma^7_{300}\left(\frac{\triangle t^{ob}}{sec}\right)}
{\xi_B^{3/2}\left(L^{ob}_{L,51}\right)^{3/2}(1+z)^4} GeV 
\,,\label{E_gamma_b_value}
\eeqarray
In the observer's reference frame on Earth the break energy in the high energy 
 photon spectrum is
\beq
E^{ob}_{{\gamma}b}=450\frac{\Gamma^8_{300}\left(\frac{\triangle t^{ob}}{sec}\right)}{\xi_B^{3/2}\left(L^{ob}_{L,51}\right)^{3/2}(1+z)^5} GeV 
\,,\label{E_g_b_ob}
\eeq
There is an acceleration time scale of the protons $``t_{acc}"$, $t_{acc}(E_p)\sim 2\pi \eta r_L(E_p)/c$ where $``r_{L}(E_p)"$ is the Larmor radius, $r_{L}(E_p)=E_p/(eB)$. $\eta$ is a factor of the order of unity. The shock accelerated protons are losing energy by synchrotron radiation of high energy photons.
 It is possible to obtain the upper cut-off energy in the proton spectrum by the condition given below
\beq
t_{acc}(E_{p, max})\simeq t_{ps}(E_{p,max})\,,\label{t_acc}
\eeq
The above condition leads us to write the following expression for the maximum proton energy in the wind rest frame.  
\beq
E_{p,max}=9.8\times 10^8\eta^{-1/2}\frac{{\Gamma_{300}}^{3/2}\left(\frac{\triangle t^{ob}}{sec}\right)^{1/2}}{\xi_B^{1/4}\left(L_{L,51}^{ob}\right)^{1/4}(1+z)}GeV \,,\label{E_p_max}
\eeq
From equation (\ref{ref_frame}) and (\ref{E_p_max}) the expression for the maximum proton energy to be observed on Earth is 
\beq
E^{ob}_{p,{\rm max}}=3\times10^{11}\eta^{-1/2}
\frac{\Gamma_{300}^{5/2}(\frac{\triangle t^{ob}}{sec})^{1/2}}
{\xi_B^{1/4}(L_{L \gamma,51}^{ob})^{1/4}(1+z)^2} GeV\,,\label{E_p_max_ob}
\eeq

The maximum photon energy to be observed on Earth can be obtained using equation
(\ref{E_p_ob}) and (\ref{E_p_max_ob}).
\beq
E^{ob}_{\gamma,{\rm max}}=1.64\times10^{4} 
\eta^{-1}(\frac{\Gamma_{300}}{1+z}) GeV\,,\label{E_g_ob}
\eeq
We have used $\eta=1$ in our numerical calculations.
For reasonable ranges of values of the GRB parameters, one observes $t_{acc}< t_{exp} \leq t_{ps}$ for $E_{p} \leq E_{pb} < E_{p,max}$, $t_{acc} \leq t_{ps} < t_{exp}$ for
$E_{pb} < E_p \leq E_{p,max}$, and $t_{ps} < t_{acc}$ for $E_p > E_{p,max}$.
 
\vskip 0.5cm
{\it The Ultrahigh Energy Proton Spectrum on Earth:}
\vskip 0.5cm

The luminosity of a GRB in very high energy photons in the observer's reference frame on Earth is
\beq
L_{H{\gamma}}^{ob}=4\pi(r_d^{s})^2 \Gamma^2 c U_{H\gamma}\,,\label{lum_high}
\eeq
, where it has been assumed that the GRB has occured at $z\approx 0$.
$r_d^{s}$ is the characteristic dissipation radius measured at the source rest frame. $U_{H {\gamma}}$ is the energy density of the very high energy photons in the wind rest frame.
If the GRB is at $z\approx0$ then the intergalactic extinction of the high energy photons due to their annihilation with infrared background photons can be neglected. If the right side of equation (\ref{lum_high}) is expressed in terms of $E_{\gamma}$, then
\beq
L_{H{\gamma}}^{ob}=4\pi(r_{d}^{s})^2 \Gamma^2 c
 \int_{E_{H{\gamma},min}}^{E_{H{\gamma},max}} E_{\gamma} \frac{dn_{\gamma}(E_{\gamma})}{dE_{\gamma}}
exp(-\tau_{in}(E_{\gamma})) dE_{\gamma}\,,\label{L_H_E_g}
\eeq
In the present calculation, $E_{H{\gamma},min}=100 GeV/{\Gamma}$, since Milagro telescope searched for photons above observed energy $100GeV$ and the upper cut-off energy of the spectrum in the wind rest frame is $E_{H{\gamma},max}=E^{ob}_{\gamma,{\rm max}}/{\Gamma}$. The internal optical depth $\tau_{in}(E_{\gamma})$ has been calculated in \cite{pijush} for various inferred values of GRB parameters. Above a few hundred GeV observed photon energy, the internal optical depth does not change with energy. It's average value between $E_{H{\gamma},max}$ and $E_{H{\gamma},min}$ can be used in the calculation. We denote it by $e_{ab}$. 
The ultrahigh energy proton spectrum in the wind rest frame and the high energy photon spectrum expected to be produced by the synchrotron radiation of these protons are related as follows
\beq
f_{\rm ps}(E_p)E_p\frac{dn_p(E_p)}{dE_p}dE_p = 
E_{\gamma}\frac{dn_{\gamma}(E_{\gamma})}{dE_{\gamma}}dE_{\gamma}\,,\label{gamma_spec1}
\eeq
 Using equation (\ref{gamma_spec1}) it is possible to write equation (\ref{L_H_E_g}) in the following form 
\beq
L_{H{\gamma}}^{ob}=4\pi(r_{d}^{s})^2 \Gamma^2 c\int_{E_{p,min}}^{E_{p,max}}
f_{\rm ps}(E_p)E_p\frac{dn_p(E_p)}{dE_p}  e_{ab} dE_p \,, \label{L_H_E_p}
\eeq
, where $E_{p,min}$ and $E_{p,max}$ can be calculated using equation (\ref{E_p_w}) and equation (\ref{E_p_max}) respectively. 
The ultrahigh energy proton spectrum expected from a GRB has two parameters, normalisation constant $A_p$ and the proton spectral index $\alpha$ which are to be determined by using the limit on high energy photon luminosity of GRBs obtained by Milagro and comparing the ultrahigh energy proton spectrum expected from GRBs with the UHECR spectrum observed on Earth by different experiments. 
Equation (\ref{L_H_E_p}) can be solved to determine $A_p$ in terms of $\alpha$,  $L^{ob}_{H{\gamma}}$, $r_d^s$, $\Gamma$, $E_{pb}$, $E_{p,min}$ and $E_{p,max}$.
\beq
A_p=\frac{L_{H{\gamma}}^{ob}}{4\pi(r_{d}^s)^2 \Gamma^2 e_{ab}c \left[\frac{E_{pb}^{(3-\alpha)}-E_{p,min}^{(3-\alpha)}}{(3-\alpha)E_{pb}} + \frac{E_{pb} (E_{p,max}^{(1-\alpha)}-E_{pb}^{(1-\alpha)})}{1-\alpha}\right]}
\eeq
 where it is assumed that $E_{p,min} < E_{pb} < E_{p,max}$.
The number of protons emitted by a single GRB per unit observed proton energy on Earth can be written as
\beq
\frac{d\phi_p(E_p^{ob})}{dE_p^{ob}}= t_d4\pi(r_{d}^{s})^2 c\frac{dn_p(E_p)}{dE_p} 
\eeq
$t_d$ is the duration of very high energy photon emission in seconds as observed from Earth. Milagro group \cite{milagro2} has used $t_d=80 sec$ for their analysis to show how an upper limit can be obtained on high energy photon luminosity of GRBs. Durations of GRBs may vary from few hundred milliseconds to tens of seconds. The shortest burst detected by BATSE (Burst And Transient Source Experiment) had a duration of $5ms$ and the longest burst so far observed had a duration of $90 minutes$. One may assume an average value for the duration of high energy photon emission from all GRBs to obtain an estimate of their contribution to the ultrahigh energy proton spectrum to be observed on Earth. The observed proton spectrum on Earth from all GRBs is expected to be
\beq
J(E_p^{ob})=\frac{1}{4\pi} R_{GRB} \frac{d\phi_p(E_p^{ob})}{dE_p^{ob}} d_c(E_p^{ob})
\eeq
, where $R_{GRB}$ is the average distribution of GRBs at $z\approx 0$.
$d_c(E_p^{ob})$ is the distance beyond which sources do not contribute to the flux above energy $E_p^{ob}$. There is discussion on the energy dependent cut-off distance $d_c(E_p^{ob})$ in \cite{waxman2,waxman3}. The spectrum of particles accelerated by Fermi mechanism in relativistic shocks has been studied in detail by several groups \cite{shock}. In the ultrarelativistic limit for a pure Kolmogorov magnetic turbulance a spectral index $\sim 2.26 \pm 0.04$ is expected \cite{martin}. We have obtained the expected ultrahigh energy proton spectrum on Earth from GRBs for $\alpha=2.2,2,1.8$.
 In particular, we have investigated about what could be the GRB parameters which may give rise to a UHECR proton spectrum on Earth comparable with the observational data from AGASA \cite{agasa1} and HiRes \cite{hires} experiments above energy $10^{19}eV$.   
\vskip 0.5cm
{\it Results and Discussions:}
\vskip 0.5cm
The ultrahigh energy proton spectra expected on Earth from all GRBs have been derived for $\alpha=2.2,2,1.8$ assuming that only one third of the protons are coming out of the GRBs and shown in Fig.1. We have used the following values of GRB parameters Lorentz factor $\Gamma=300$, low energy photon luminosity $L^{ob}_{L,{\gamma}}=10^{51} erg/sec$, intrinsic variability time of the source as measured on Earth $\triangle t^{ob}=5sec$, high energy photon luminosity $L^{ob}_{H,{\gamma}}=10^{51} erg/sec$ and distribution of GRBs $R_{GRB}=4.8 GRBs/Gpc^3/yr$ at $z\approx 0$. We have considered the low energy photon (KeV-MeV) spectrum from GRBs with its typical value of break energy $\epsilon^{ob}_{b}=0.5 MeV$. The spectrum of low energy photons is a power law spectrum in energy with spectral indices $\beta_l=1$ and $\beta_h=2.25$ below and above the break energy respectively.
This low energy photon spectrum has been used to calculate the internal optical depth of high energy photons. The internal optical depth has a value $0.098$ at $100 GeV$ and its value is $0.196$ at $1000GeV$ for $\Gamma=300$, $L^{ob}_{L,{\gamma}}=10^{51}erg/sec$, $\triangle t^{ob}=5 sec$, $z=0$, $\epsilon^{ob}_b=0.5 MeV$ and $\beta_l=1,\beta_h=2.25$. We have used the average value of the internal optical depth between observed photon energies $100 GeV$ and $1000 GeV$ in our calculations. The value of the internal optical depth almost saturates to a constant value at higher energies \cite{pijush}. The particle flux decreases rapidly with increasing energies hence, a slight variation in the value of $\tau_{in}(E_{\gamma})$ with increasing energies would not affect our results although we are using the exponential of $\tau_{in}(E_{\gamma})$ in our calculations.  
The break energy, lower and upper cut-off energies in the proton spectrum in the wind rest frame are $8\times10^8 GeV$, $1.67\times 10^8 GeV$ and $21.9\times10^8 GeV$ respectively. The ultrahigh energy proton spectrum obtained by us for $\alpha=1.8$ in Fig.1. is comparable with the data recorded by HiRes experiment but below the data points of AGASA experiment. Below the peak the spectrum with spectral index $\alpha=2.2$ is comparable with the data collected by AGASA experiment but it is above the data points recorded by HiRes experiment.
\para

There are several parameters which are involved in this study. It is important to dicuss how our results depend on the parameters like the observed variability time of GRBs $\triangle t^{ob} sec$, the Lorentz factor $\Gamma$, and the duration of very high energy photon emission from GRBs $t_d sec$. The internal optical depth $(\tau_{in}(E_{\gamma}))$ of high energy photons is inveresly proportional to $\triangle t^{ob}$ and $\Gamma^6$ or $\Gamma^4$ depending on the photon energy \cite{pijush}. It is directly proportional to the observed luminosity of low energy photons (KeV-MeV) $L^{ob}_{L,{\gamma}}$. 

\para

If $\triangle t^{ob}=0.01 sec$ instead of $5sec$, then the value of the internal optical depth increases by a factor of 500 and no high energy photons would be able to come out from GRBs. An increase in the the value of $\triangle t^{ob}$ to $10 sec$ from $5 sec$ does not give rise to any change in the high energy proton or photon spectrum from GRBs. 

\para

Internal optical depth is very sensitive to the changes in the value of the Lorentz factor $\Gamma$. If $\Gamma \sim 1000$ instead of $300$ and the values of the other parameters remain unchanged, then the value of the internal optical depth becomes $123$ times smaller hence, all high energy photons would be able to come out of GRBs. The break energy $(E_{pb})$ and the upper cut-off energy $(E_{p,max})$ in the ultrahigh energy proton spectrum from GRBs are proportional to $\Gamma^5$ and $\Gamma^{3/2}$ respectively. For $\Gamma=1000$, the break energy $E_{pb}$
 becomes $\sim 3.29\times 10^{11} GeV$ and it is higher than the maximum energy in the proton spectrum $E_{p,max}=1.33\times 10^{10} GeV$.  
In this case the expected ultrahigh energy proton flux on Earth from all GRBs would be one order of magnitude higher than the observational data of HiRes experiment even after assuming an efficiency of proton emission equal to $1/3$.
If the Lorentz factor is 100 then the internal optical depth would be $81$ times larger than the corresponding value for $\Gamma=300$ hence, very high energy photons would not be able to come out of the GRBs. The ultrahigh energy proton flux on Earth would be very high for $\Gamma=100$, if we assume that the high energy photon luminosity of GRBs is $10^{51} erg/sec$ as obtained by Milagro group for $R_{GRB}=4.8 GRBs/Gpc^3/yr$ and $t_d=80 sec$.

\para
 
If the total energy emitted in very high energy photon emission by each GRB as measured on Earth is denoted by $E^{ob}_{H{\gamma}}$ then, $E^{ob}_{H{\gamma}}=L^{ob}_{H{\gamma}} \times t_d$. The expected ultrahigh energy proton flux on Earth from all GRBs is directly proportional to $E^{ob}_{H{\gamma}}$. An increase in the value of $L^{ob}_{H{\gamma}}$ or $t_d$ would result in an enhancement of the ultrahigh energy proton flux on Earth to be produced from all GRBs. 

\para

 Milagro group has assumed the high energy photon spectrum to be proportional to $E^{-2}_{\gamma}$ in deriving the upper limits on high energy photon luminosity of GRBs. We have used their limit in the context of the proton synchrotron model of high energy photon production. In this model the photon spectrum is proportional to $E^{-(\alpha+1)/2}_{\gamma}$ and $E^{-(\alpha+2)/2}_{\gamma}$ below and above the break energy respectively where, $\alpha \sim 2.$ \cite{totani,pijush}. The discrepancy in the values of the high energy photon spectral index used in their analysis and the present work has to be kept in mind while explaining
 our results. Also, the values of the GRB parameters are not known to us at the time of their very high energy photon emission. It may be possible that different GRBs are emitting different energies in very high energy photon emission. 
 We do not claim our results to be accurate as there are several uncertainities involved in these calculations. We have derived what could be the ultrahigh energy proton spectrum on Earth  from all GRBs if we use the upper limits on their luminosity of very high energy photon emission obtained by using high energy $\gamma$-ray telescope. 
\vskip 0.5cm
{\it Conclusion:}
\vskip 0.5cm
 The ultrahigh energy proton spectra on Earth from all GRBs have been derived for $\alpha=2.2,2,1.8$ assuming  Lorentz factor to be 300, isotropic low energy photon luminsoity of GRBs as observed on Earth to be $10^{51} erg/sec$, variability time of GRBs to be $5 sec$ as measured on Earth, observed isotropic high energy photon luminosity of GRBs to be $10^{51} erg/sec$, duration of high energy photon emission to be $80 sec$ and assuming that GRBs are isotropically distributed with a rate of $4.8 GRBs/Gpc^3/yr$ near $z\approx0$. If the efficiency of proton emission from GRBs is assumed to be equal to $1/3$ then our derived spectrum for $\alpha=1.8$ becomes comparable with the data collected by HiRes experiment. 
\vskip 0.5cm
{\it Acknowledgment:}
\vskip 0.5cm
The author would like to thank M. F. Morales and P. Bhattacharjee for helpful communications.

\newpage 
\begin{figure}
\begin{center}
\centerline{
\epsfxsize=20.cm\epsfysize=10.cm
\epsfbox{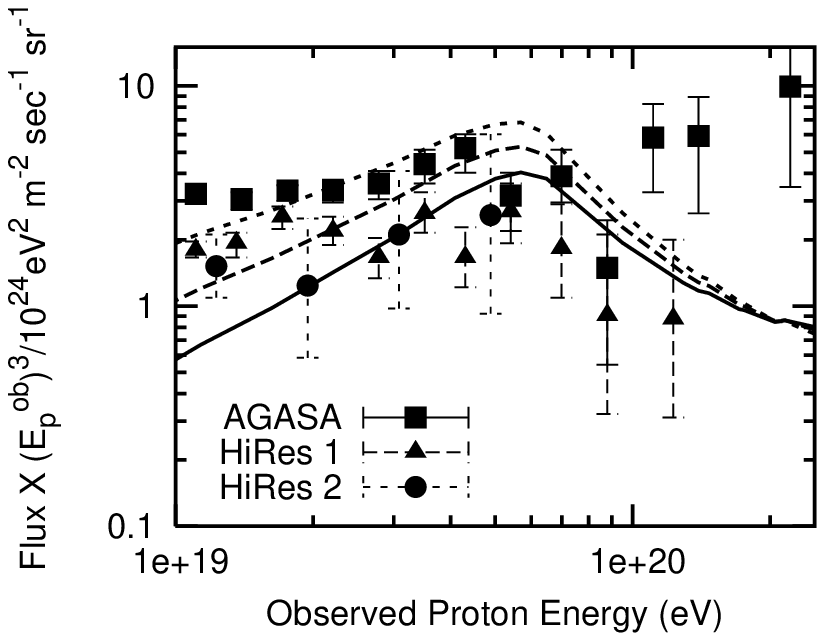} 
}

\end{center}
\caption{Proton spectra on Earth from all GRBs have been derived with $\alpha=2.2,2,1.8$ (dotted, dashed, solid lines) and compared with the observational data from AGASA \cite{agasa1} and HiRes \cite{hires} experiments. The values of the GRB parameters used to derive the UHECR proton spectra on Earth from all GRB are as follows: $\Gamma=300$, $L^{ob}_{L{\gamma}}=10^{51}erg/sec$, $\triangle t^{ob}=5 sec$, $L^{ob}_{H,{\gamma}}=10^{51}erg/sec$ and $t_d=80 sec$.}
\end{figure}
\end{document}